\documentclass[12pt]{article}
\usepackage{graphicx} 
\usepackage[top=0.85in,right=1in,left=1in,footskip=0.75in,marginparwidth=1in]{geometry}
\usepackage{amsmath, amsfonts, amssymb, bm}
\usepackage{placeins}
\usepackage{natbib}
\usepackage{url}
\usepackage[colorlinks=true,linkcolor=blue, citecolor=blue]{hyperref}
\usepackage{longtable,booktabs,multirow,array}
\newcolumntype{P}[1]{>{\raggedright\arraybackslash}p{#1}}

\title{The Multidimensional Index of Child Growth (MICG) of the Task Force ``Towards a Multidimensional Approach for Child Growth'' of the International Union for Nutrition Sciences}

\date{University of Groningen, The Netherlands. November 2025}

\author{
  Rolando Gonzales Martinez\\
  \texttt{r.m.gonzales.martinez@rug.nl}
  \and
  Hinke Haisma\\
  \texttt{h.h.haisma@rug.nl}
}

\begin{document}

\maketitle

\begin{abstract}
\noindent Children’s growth is more than height and weight. In this document we introduce and explain the Multidimensional Index of Child Growth (MICG), developed by the International Union for Nutrition Sciences (IUNS) Task Force ``Towards a Multidimensional Approach for Child Growth,'' to operationalize a capability- and rights-based view of child development that spans 14 physical and non-physical dimensions (e.g., life and physical health, love and care, mental wellbeing, participation, time autonomy, mobility, safety). We prototyped our approach, using the Young Lives Study for Ethiopia, India (Andhra Pradesh and Telangana), Peru, and Vietnam. As a proof-of-concept, we operationalized the theoretical multidimensional framework with 29 indicators grouped into 14 dimensions, and computed MICG via a counting approach using the Young Lives Survey. We compare endogenous (data-driven) and exogenous (equal and child-prioritized) weighting schemes, finding high rank concordance and demonstrating that equal weights are a robust, policy-pragmatic choice. MICG reveals deprivation patterns obscured by anthropometrics alone—e.g., rural girls in Peru show compounded shortfalls in education, mobility, and mental wellbeing despite similar physical status to peers. Quantile models and regression analyses highlight how community participation in the design of WASH programs is associated with higher multidimensional achievements, especially among the most deprived, underscoring the value of inclusive programming and planning. To bridge the gap between achievements and unrealized potential, we extend MICG with a Bayesian stochastic-frontier formulation that estimates opportunity distributions (capabilities) conditional on context (sex, geography), pinpointing children likely to be “left behind” even when current outcomes appear similar. We also propose a spiderweb growth chart for monitoring multidimensional achievements of child growth at both individual and population levels, complementing WHO anthropometric charts and supporting clinical, programmatic, and cross-country comparisons aligned with the Sustainable Development Goals (SDGs). MICG offers an actionable, equity-sensitive diagnostic tool for monitoring and evaluating nutrition-specific and nutrition-sensitive interventions, strengthening surveillance, and advancing the SDG pledge to leave no child behind.
\end{abstract}

\section{Introduction}

Over the past decades, Ministries of Health, NGOs, and scientists have made substantial progress in reducing child mortality around the globe \citep{UNIGME2023,You2022Lancet}. However, many children still do not reach the age of five \citep{UNIGME2025,GBDChildMortality2024}. The double and triple burden of malnutrition creates new challenges for healthy child growth, particularly in low-resource settings \citep{Popkin2020,Victora2021}. Scientists and policymakers agree on the importance of multisectoral programmes to combat malnutrition and improve child morbidity and mortality \citep{Ruel2013,Escher2024}. 

To achieve this goal, the design, implementation, and assessment of nutrition-specific and nutrition-sensitive interventions call for a multidimensional approach to child growth that goes beyond physical outcomes. A multidimensional assessment of child growth—aimed to understand child development beyond anthropometry—sheds new light on inequalities and helps design interventions better tailored to the needs of vulnerable children \citep{Black2021}. The rights-based multidimensional framework contributes to the SDG agenda of leaving no one behind and could be extended to clinical practice for growth monitoring and assessment \citep{UNICEF2020}.

The Task Force ``Towards a Multidimensional Index for Child Growth'' of the International Union for Nutrition Sciences (IUNS) was launched at the International Conference on Nutrition in 2013. This project was initiated in 2012 and received funding from NWO/WOTRO/VIDI (W01.70.300.002) until 2018. In a workshop attended by experts from different disciplines, multidimensional child growth was conceptualized as ``the process of continuous physical, psychological, and social change that builds a child's capacities to maximize life chances at the individual and societal level'' \citep{Haisma2018}. This concept is grounded in theories from diverse disciplines, such as nutrition transition theory \citep{Popkin1993,Popkin2004}, parent–offspring conflict theory \citep{Trivers1974}, and life-history theory from evolutionary biology \citep{Stearns1992}. Building on Biggeri's conceptual model for child wellbeing \citep{Biggeri2001,BiggeriCuesta2021} and Bronfenbrenner's ecological model \citep{Bronfenbrenner1995}, the Task Force developed the Capability Framework for Child Growth. Following Nussbaum and Biggeri, a rights-based approach was adopted to identify the dimensions of child growth \citep{Nussbaum2011,Yousefzadeh2019}.

The IUNS–MICG project started with a conceptualization of healthy child growth embedded in the capability approach and the \textit{Convention on the Rights of the Child} \citep{UNCRC1989}. After the conceptualization of multidimensional child growth, in Expo 2015 in Milan (a project created in collaboration with the Giangiacomo Feltrinelli Foundation), scientists were invited to discuss the need and potential of a multidimensional approach to child growth. Since then, the Task Force has provided a platform for scientists to contribute to the development of the approach. In the first term (2013–2017), the focus was on theoretical foundations and conceptualization. In the second term (2017–2022), the focus shifted to operationalization and contextualization through empirical research in Tanzania and Bangladesh \citep{
Chakraborty2022,
Chakraborty2020a,
Chakraborty2020b,
Mchome2019a,
Mchome2020a,
Mchome2020b,
Mchome2020c,
Mchome2021}. In the third term (from 2022 onward), a multidimensional index was generated to assess interventions and enable cross-country comparisons in a multidimensional way. During this third therm, a methodology for calculating The Multidimensional Index of Child Growth (MICG) was developed to identify children who are deprived in various dimensions of development, using the \textit{Young Lives} database \citep{YoungLives2022}. After coding twenty-nine indicators of the Young Lives Survey, these were grouped under their dimensions. Within each dimension, indicators were equally weighted; each dimension then received equal weight in the overall MICG to reflect parity between physical and non-physical aspects. A child was identified as multidimensionally deprived if the weighted deprivation score crossed a pre-specified threshold (e.g., $k = 1/3$ of total weighted deprivations), following established practice in multidimensional poverty measurement \citep{AlkireFoster2011,AlkireFoster2016}. This approach moves beyond a narrow reliance on physical growth indicators to capture multiple, intersecting deprivations that shape child development. Finally, a deprivation score $D_i$ is calculated as the weighted sum of indicators for child $i$, and the MICG achievement $A_i$ is defined as $A_i = 1 - D_i$. We consider both \emph{endogenous} and \emph{exogenous} weights: endogenous weights derived statistically from the covariance structure of the data (e.g., via principal components or factor models), and exogenous weights based on normative judgments about the relative importance of dimensions \citep{DecancqLugo2013}. Robustness was assessed by comparing rankings across weighting schemes. The children identified as deprived by the IUNS-MICG methodology are the ones at risk of being left behind by development policies in multiple dimensions (biological, psychological, and social) and would benefit from tailored interventions targeting specific domains of deprivation, such as economic exploitation or exposure to domestic violence. The MICG approach is relevant to individual, population, and health-system analysis at micro- (individual, household), meso- (regional), and macro- (national) levels. 

Section~\ref{sec:capability} describes the capability approach for children. Section~\ref{sec:operationalization} explains the operationalization of the capability approach—translating the theoretical framework into an empirical index of multidimensional child growth. Section~\ref{sec:applications} shows how MICG can be applied at macro-, meso-, and micro-levels. Section~\ref{sec:conclusions} concludes. Appendix A shows how a multidimensional approach helps identify the worst-off children using Demographic and Health Survey (DHS) and Young Lives data. Appendix B provides a non-Euclidean Extension of Bronfenbrenner’s Bioecological Theory in which chidl growth is conceptualized as motion across a curved ecological manifold, not movement through fixed layers. Inequality, resilience, and adaptation appear as geometric properties---asymmetries, curvatures, and torsions---of the social universe that surrounds children.

\section{Conceptual approaches and the capability framework}\label{sec:capability}

Several approaches and methodologies have been suggested to conceptualize and measure child well-being and child development. These approaches are rooted in different scientific disciplines, including nutrition science \citep{Requejo2022}, philosophy \citep{Hassoun2024}, psychology \citep{Lippman2011}, and economics/policy measurement \citep{OECD2021}. Examples of widely used indices and frameworks include the WHO-led Global Scales for Early Development (GSED) that deliver a common developmental score (the D-score) \citep{WHO_GSED_2023,McCray2023}, UNICEF's Early Childhood Development Index 2030 (ECDI2030) \citep{UNICEF_ECDI2030_2023,UNICEF_ECDI2030_Guidance_2023}, the Nurturing Care Framework \citep{Wertlieb2019}, and the Child Flourishing perspective articulated by the WHO--UNICEF--Lancet Commission \citep{Clark2020}.

Within nutrition science, child growth has often been conceptualized narrowly as change in weight and height, i.e., reflecting primarily the physical dimension. The IUNS Task Force advocates re-defining child growth within the capability approach \citep{Haisma2018,Yousefzadeh2019}, rendering it a multidimensional concept that explicitly includes the environmental context in which a child grows, since this context conditions physical, social, and psychological development. Capabilities capture not only a child's attained state but also the \emph{opportunities} to grow in physical and non-physical dimensions, conditional on the situation of parents and caregivers.

The capability approach guides researchers in building multidimensional indices. In capability theory, \emph{endowments} are resources or assets a person possesses (e.g., tangible goods, personal traits, social entitlements). These can be converted into \emph{functionings}---achieved outcomes such as being healthy, safe, educated, or mobile. The conversion from endowments to functionings depends on \emph{conversion factors}, which may be personal (ability, knowledge), social (norms, discrimination), or environmental (infrastructure, climate). Following Sen's original formulation \citep{Sen1999}, people choose functionings by converting commodities into vectors of characteristics (e.g., a bicycle can provide the characteristics of play and transportation). In a child-focused capability approach, two features are central: (1) capabilities are age-dependent; and (2) there is an intergenerational transfer of capabilities from parents and caregivers to children. For example, recent evidence shows that maternal anaemia at 28 weeks of pregnancy is associated with an increased risk that children do not attain expected educational standards at age five \citep{Olga2024}.

Thus, for children, capability theory must be adjusted to account for parental/caregiver endowments and agency. The child's achievements (e.g., being well-nourished) are \emph{functionings}, and \emph{capabilities} are the opportunities to reach these functionings (e.g., the opportunity to have a nutrient-diverse meal). Children's conversion factors are influenced by those of their parents and caregivers.

Formally, let the capability set of child $i$ be $Q_i(E_i)$, defined as the set of attainable functionings $a_i$ that result from applying a choice function $f_i \in F_i$ to the outcome of a conversion function $c(\cdot)$ acting on the child's endowments $e_i \in E_i$:
\begin{equation}
  Q_i(E_i) \;=\; \big\{\, a_i \;=\; f_i\!\big(c(e_i)\big) \ : \ f_i \in F_i,\ e_i \in E_i \,\big\}.
\end{equation}
Here, $Q_i(E_i)$ is the capability set; $a_i$ is the achieved functioning; $f_i$ represents the child's choice; $c(\cdot)$ is the conversion function; and $E_i$ is the set of endowments available to the child. This basic structure assumes a direct transformation from endowments to functionings through conversion and choice.

For children, this framework must be expanded to reflect deep dependence on parents/caregivers. Let parental endowments be $E_p$ with conversion function $c_p(\cdot)$. Parental endowments $e_p \in E_p$ are transformed into enabling conditions for the child. The child's conversion function $c_i(\cdot)$ now takes two arguments---the child's endowments $e_i$ and the converted parental conditions $c_p(e_p)$. The resulting achieved functioning $a_i$ is:
\begin{equation}
  Q_i(E_i, E_p) \;=\; \big\{\, a_i \;=\; f_i\!\big(c_i\big(e_i,\ c_p(e_p)\big)\big) \ : \ f_i \in F_i,\ e_i \in E_i,\ e_p \in E_p \,\big\}.
\end{equation}
This formulation acknowledges a layered structure in which parental capacities, resources, agency, and choices directly shape the child's context. The child's conversion factors---and ultimately their functionings---are embedded within, and shaped by, prior parental conversions. This recursive logic aligns with empirical observations that children's wellbeing and opportunities are fundamentally mediated by caregiver conditions.

Moreover, the capability approach includes the \emph{agency} of children---their voice, choice, and participation in matters affecting their lives (education, play, relationships, cultural expression). This shifts policy from a purely top-down model (what adults think children need) to one that includes children's perspectives (e.g., participatory education methods, child-inclusive surveys, rights-based approaches).

Functionings in multidimensional child growth are also age-dependent: the same endowment can enable different functionings across ages and contexts. For instance, a bicycle may support leisure (play) or efficient commuting to school; but if parents cannot teach the child to ride, roads are unsafe, or social norms discourage use, conversion factors hinder the functioning. Similarly, access to vaccination depends on trust, time, logistics, and services---conversion factors for both children and caregivers.

\begin{figure}
\centering
\includegraphics[width=\textwidth]{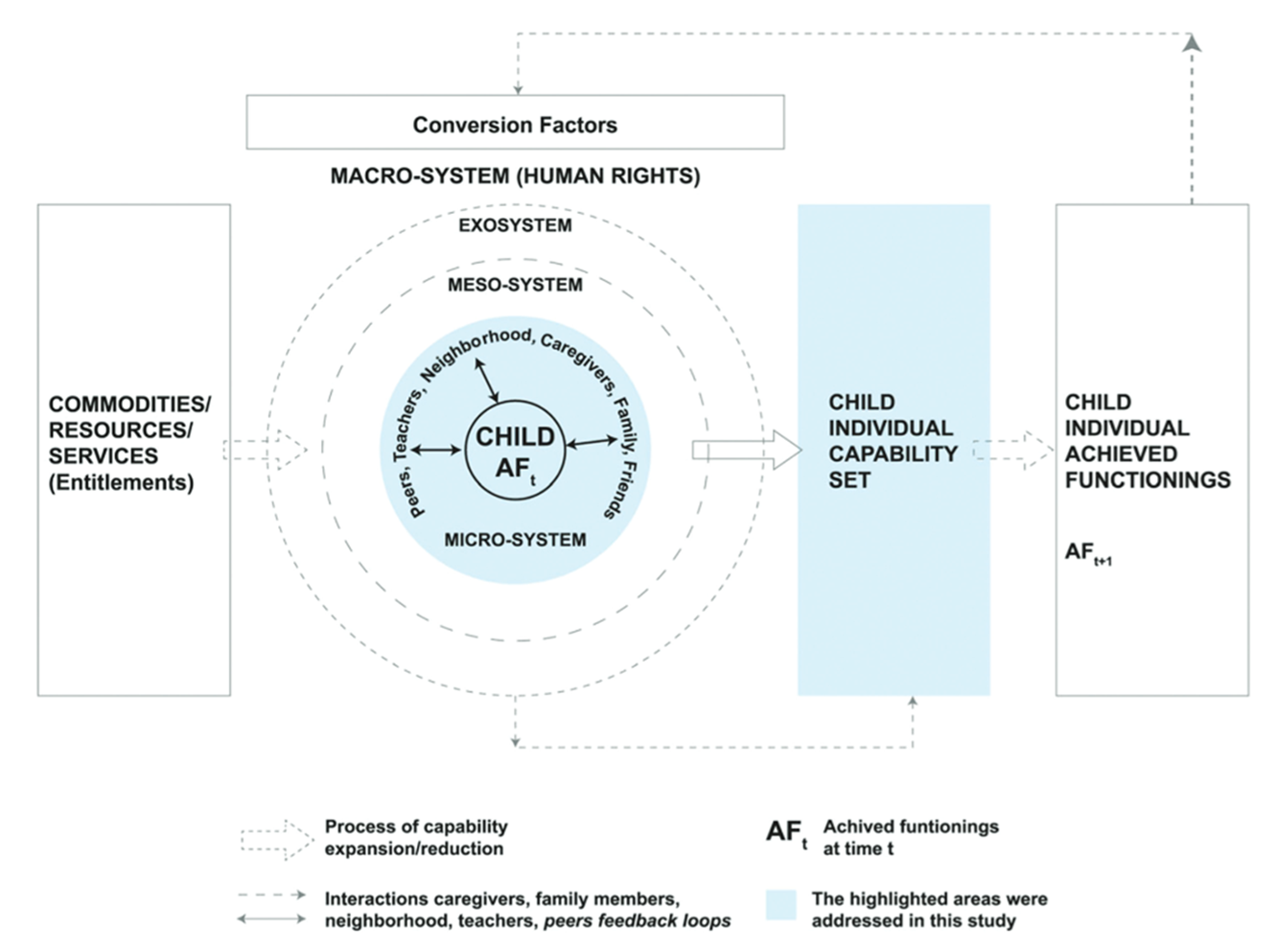}
\caption{Conceptual framework of the Multidimensional Index of Child Growth (MICG) linking capabilities, functionings, and conversion factors across bio-ecological levels.}
\label{fig:framework}
\end{figure}

Finally, Figure \ref{fig:framework} sketches a multidimensional capability framework for child growth, in which physical and non-physical development emerges from interactions across ecological systems. At the center is the child, whose achieved functionings reflect current wellbeing; these are influenced by micro-systems (family, caregivers, teachers, peers) nested within meso-, exo-, and macro-systems (neighborhoods, institutions, and human-rights frameworks).

\section{Operationalizing the capability approach for children}\label{sec:operationalization}

Operationalizing the capability approach for children implies translating the theoretical framework of the TMICG--IUNS into an empirical index. This is challenging because of uncertainty about which dimensions to select and because capabilities (opportunities) are not directly observable. The \emph{unobservability} of capabilities implies that statistical methods are needed to estimate the real opportunities that children have across multiple, inter-related dimensions \citep{AlkireFoster2011,Robeyns2017}.

An extensive literature indicates which dimensions are relevant for a MICG. \citet{Nussbaum2011} identifies core physical and non-physical capabilities for human flourishing, including life, bodily health, bodily integrity; senses, imagination, and thought; emotions; practical reason; affiliation; play; and control over one's environment. In children, synergies between physical and non-physical dimensions imply, for example, that a child unable to attend school due to malnutrition faces reduced opportunities to acquire education.

Building on the \textit{Convention on the Rights of the Child} and participatory work with children, \citet{Biggeri2006} proposed fourteen dimensions for child wellbeing: life and physical health; bodily integrity and safety; love and care; leisure; respect; social relations; participation; mental well-being; education; freedom from economic and non-economic exploitation; environment (internal and external risks); religion and identity; time autonomy; and ease of mobility to education or health centers. These fourteen dimensions were discussed and prioritized by children during the first Children’s World Congress on Child Labor (Florence, May 2004), with balanced participation of girls and boys aged 10--17 selected through national and regional consultations intended to minimize discrimination \citep{Biggeri2006}. While not designed to be statistically representative of all children, the sample was selective and of high quality due to participants' prior engagement with NGOs and lived experience as former child laborers and youth activists, which informed their perspectives on child needs.

For the operationalization of MICG as an index, we include these fourteen dimensions. Data on indicators relevant to multidimensional child growth for each dimension must be obtained either by primary data collection or by selecting variables from a suitable database. Because children’s opportunities and achievements depend on parental attitudes and on the surrounding environment, MICG includes indicators that capture caregiver characteristics and contextual conditions \citep{Haisma2018,Yousefzadeh2019}.

For a population-level prototype, we identified \textit{Young Lives} (YL) as sufficiently rich in variables capturing opportunities across the fourteen dimensions. YL follows two cohorts per country: a younger cohort of about 2{,}000 children (6--18 months in 2002) and an older cohort of about 1{,}000 children (7.5--8.5 years in 2002). Round~2 provides cross-sectional information when the younger cohort reached age five across sentinel sites in Ethiopia, Peru, Viet Nam, and the Indian states of Andhra Pradesh and Telangana, including contextual information on children’s environments and parental attitudes \citep{YL_SamplingGuide2015,YL_Ethiopia_R2,YL_Peru_R2,YL_VN_Factsheet}. Round~2 instruments comprised a child questionnaire (age 5), a household questionnaire for parents/caregivers, and a community questionnaire for local stakeholders. The child/household instruments cover parental background, education, livelihoods, consumption, social capital, socioeconomic status, childcare, education, health, anthropometry, caregiver attitudes, school activities, social networks, feelings and attitudes, and future perceptions; the community instrument covers locality characteristics, social environment, service access, and modules on education and health services. YL’s sentinel-site design balances regional diversity and rural/urban differences in heterogeneous agro-climatic areas with a pro-poor focus; multi-stage purposive sampling was used to obtain deep insight into groups of interest where phenomena are unevenly distributed \citep{YL_SamplingGuide2015}.

Table \ref{tab:micg_yls_indicators} lists the twenty-nine indicators used to measure the fourteen MICG dimensions and the cut-offs that define a child as deprived on each indicator. Relevant questions from YL were mapped to MICG dimensions. For \textit{Life and physical health}, anthropometry (height, weight, age) informed stunting, wasting, and overweight indicators; dietary recall and food-group consumption informed nutrition frequency and diversity. Household items on shelter quality, access to education, time spent on domestic chores, and caregiver attitudes mapped to \textit{bodily integrity}, \textit{education}, \textit{time autonomy}, and \textit{love and care}. Non-physical indicators captured caregiver attitudes and practices, cognitive and verbal development, domestic tasks, and internal/external safety risks.

Cut-offs to identify deprivations followed international norms and technical guidance and were refined through expert consultation in nutrition and non-physical child development. For example, a child is considered stunted if height-for-age $z$-score $< -2$ SD, wasted if weight-for-length/height $z$-score $< -2$ SD, and deprived in education if schooling is not accessible. A child is deprived in safety if caregivers report that it is unsafe to go to school or play in the street. Anthropometric standards followed WHO growth standards \citep{WHO2006standards} and the WHO--UNICEF joint statement on identification of severe acute malnutrition \citep{WHO2009sam}.

% Preamble (once in your main .tex)
% \usepackage{longtable,booktabs,multirow,array}
% \newcolumntype{P}[1]{>{\raggedright\arraybackslash}p{#1}}

\begin{longtable}{P{3cm} P{3.6cm} P{8cm}}
\caption{Indicators included in the multidimensional index of child growth (MICG) based on the availability of information from the Young Lives Survey (Round 2)}\label{tab:micg_yls_indicators}\\
\toprule
\textbf{Capability} & \textbf{Indicator} & \textbf{Deprivation cut-off for the indicator} \\
\midrule
\endfirsthead

\toprule
\textbf{Capability} & \textbf{Indicator} & \textbf{Deprivation cut-off for the indicator} \\
\midrule
\endhead

\midrule
\multicolumn{3}{r}{\emph{Continued on next page}}\\
\bottomrule
\endfoot

\bottomrule
%\multicolumn{3}{l}{\footnotesize Notes: Cut-offs follow international standards (e.g., $\pm$2 SD) and expert consultations in nutrition and non-physical child growth.}\\
\endlastfoot

\multirow{7}{=}{Life and physical health}
  & overweight             & Child deprived if BMI $>$ 2 SD (WHO thresholds by age/sex). \\
  & stunting               & Deprived (stunted) if length/height-for-age $z$-score $< -2$ SD. \\
  & wasting                & Deprived (wasted) if weight-for-length/height $z$-score $< -2$ SD. \\
  & nutrition (frequency)  & Deprived if child eats $<$ 4 times per day. \\
  & nutrition (diversity)  & Deprived if child eats $<$ 4 different food groups per day. \\
  & vaccination            & Deprived if child does not have a vaccination card. \\
  & health                 & Deprived if child is reported to have worse health than peers. \\
\midrule \addlinespace

\multirow{3}{=}{Bodily integrity and safety}
  & safety in the street   & Deprived if parents think it is not safe for the child to go on the street on their own. \\
  & shelter                & Deprived if household dwelling has rudimentary roof/wall materials (straw, wood, leaves, mud, plastic sheets). \\
  & danger                 & Deprived if child feels/would feel in danger when travelling to school. \\
\midrule \addlinespace

\multirow{4}{=}{Love and care}
  & love of parents        & Deprived if parents believe parent--child love is not important at all / not very important. \\
  & proud of children      & Deprived if parents \emph{strongly disagree} with feeling proud of their children. \\
  & parents' responsibility& Deprived if parents believe a sense of responsibility is not important at all / not very important. \\
  & parents' fulfillment   & Deprived if the pleasure parents get from watching children grow is not important at all / not very important. \\
\midrule \addlinespace

Leisure activities
  & leisure                & Deprived if child does not spend time with friends/younger children, playing games, watching TV, playing alone or with pets. \\
\midrule \addlinespace

Respect
  & respect                & Deprived if parents believe it is not important to learn (i) responsibility, (ii) obedience, or (iii) respect for elders. \\
\midrule\addlinespace

Social relations
  & friendship             & Deprived if child has difficulties making friends or cannot make friends at all. \\
\midrule\addlinespace

Participation
  & cooperation            & Deprived if parents think it is not important to learn cooperation/participation at home. \\
\midrule\addlinespace

\multirow{2}{=}{Mental well-being}
  & cognitive              & Deprived if Rasch cognitive development score is below the first quintile. \\
  & verbal                 & Deprived if Rasch verbal PPVT score is below the first quintile. \\
\midrule\addlinespace

\multirow{2}{=}{Education}
  & access to education    & Deprived if the household does not have access to education. \\
  & imagination            & Deprived if parents believe it is not important to learn imagination at home. \\
\midrule\addlinespace

Economic freedom
  & paid/unpaid work       & Deprived if child performs paid activities or works on family farm/business on a typical day. \\
\midrule\addlinespace

\multirow{2}{=}{Environment}
  & external risks         & Deprived if child is exposed to natural hazards or harassment from other children, authorities, or rebels/thieves. \\
  & internal risks         & Deprived if number of prior births to the same mother exceeds caregivers' stated ideal number. \\
\midrule\addlinespace

\multirow{2}{=}{Religion and identity}
  & children's religion    & Deprived if child has a minority religion. \\
  & children's ethnicity   & Deprived if child belongs to an ethnic minority. \\
\midrule\addlinespace

Time autonomy
  & hours of domestic tasks& Deprived if child cares for others or performs domestic tasks on a typical day (above threshold). \\
\midrule\addlinespace

Mobility
  & time to get to school  & Deprived if child needs/would need more than one hour to get to school. \\

\end{longtable}

The 29 indicators were aggregated into the Multidimensional Index of Child Growth (MICG) following a counting approach inspired by the Alkire–Foster methodology \citep{AlkireFoster2011, AlkireFoster2016}. Each indicator was first coded as deprived (1) or non-deprived (0) using the cut-offs described in Table~1. Indicators were then grouped under their corresponding 14 dimensions, and within each dimension, indicators were equally weighted so that the dimension carried the same importance regardless of the number of indicators it contained. Subsequently, each dimension was assigned an equal weight in the overall MICG, reflecting the multidimensional nature of child growth and ensuring that non-physical aspects such as love and care or social participation contributed on par with traditional anthropometric measures. 

A child was classified as multidimensionally deprived if their weighted deprivation score crossed a specified cut-off (e.g., one-third of the total weighted deprivations), consistent with international practice in multidimensional poverty and child well-being research. This procedure allowed us to move beyond a narrow reliance on physical growth indicators and instead capture the multiple, intersecting deprivations that shape children’s development.

After identifying all deprivations, a deprivation score is calculated as the weighted sum of indicators. The MICG is then expressed as one minus this weighted deprivation score, representing the child’s overall achievements. The weights for each indicator can be determined either endogenously or exogenously. Endogenous weights are derived statistically from the covariance structure of the data, while exogenous weights are based on normative judgments regarding the relative importance of each dimension.

\begin{figure}
\caption{Kernel density of the MICG for Ethiopia with different weighting schemes.}
\medskip
\centering
\includegraphics[width=0.7\textwidth]{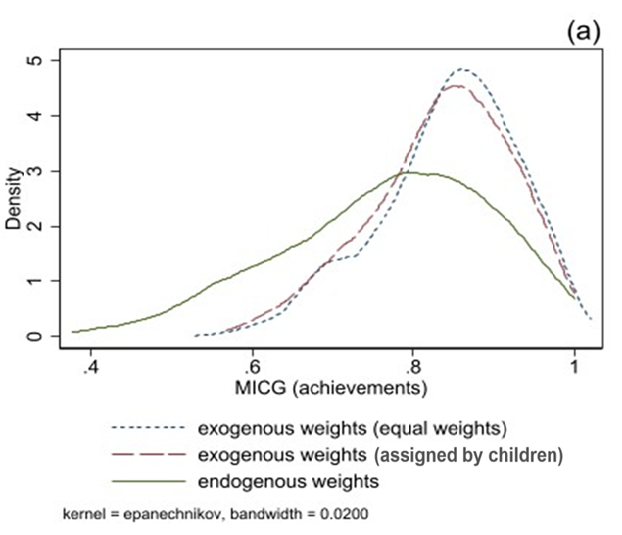}
\label{fig:den}
\end{figure}

A comparative analysis was performed to evaluate how sensitive the results are to the use of endogenous versus exogenous weights. Endogenous weights were obtained through binary principal component analysis, while exogenous weights were chosen using two weighting schemes: equal weights and weights based on the preferences of children. When children were asked what the most important opportunities in life are, they prioritized education and love and care over physical health. Figure \ref{fig:den} shows the kernel density of the MICG of Ethiopia obtained with the three different types of weights. Endogenous weights calculated with data produced a more dispersed index of child growth compared to exogenous weights.

\begin{figure}
\caption{Correlation matrix between the MICG estimates for Ethiopia with different weights.}
\medskip
\centering
\includegraphics[width=0.7\textwidth]{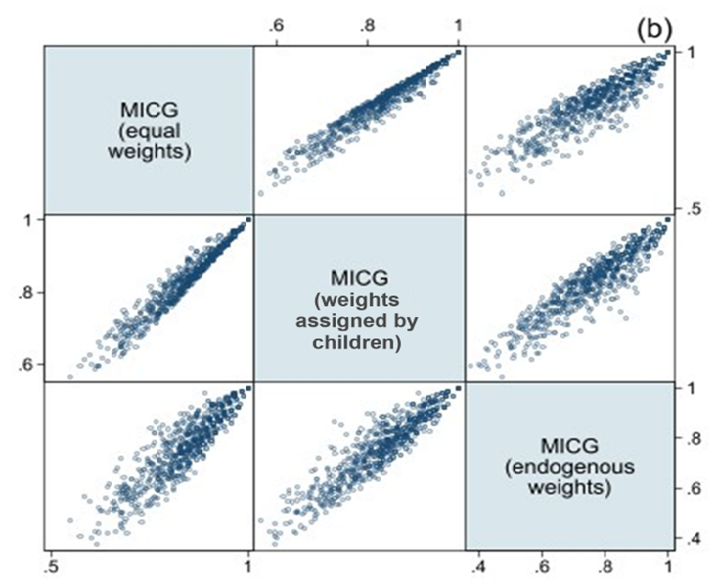}
\label{fig:comat}
\end{figure}

Figure \ref{fig:comat} shows the correlation between the MICG estimates. The MICG calculated with endogenous weights is strongly correlated with the MICG calculated with exogenous weights chosen through a children’s participatory process (Spearman’s rank correlation = 0.91, $p<0.0001$), but the highest correlation (Spearman’s rank correlation = 0.97, $p<0.0001$) is obtained between the MICG calculated with equal weights and the MICG calculated with exogenous weights chosen by children. Consequently, applying equal weights is a robust strategy with minimal impact on the calculation of the aggregated index of child growth.

Additionally, since weights derived from statistical methods have been criticized for violating key properties of well-being indices such as monotonicity and subgroup consistency \citep{dutta2021endogenous}, we assign equal (non-statistical) weights to each dimension in the results section. This choice is further justified by the fact that similar results were obtained when comparing endogenous weights derived from the data and exogenous weights informed by children’s perspectives. Because the 14 dimensions of the MICG are aligned with the United Nations Convention on the Rights of the Child (CRC), equal weighting reflects the principle that all children’s capabilities and rights are equally important. Consequently, dimensions with more indicators distribute their weight across several measures, while dimensions with fewer indicators allocate a larger share of weight to each indicator.

\begin{figure}
\caption{Spiderweb: multidimensional chart based on physical and non-physical dimensions of child growth}
\medskip
\centering
\includegraphics[width=0.7\textwidth]{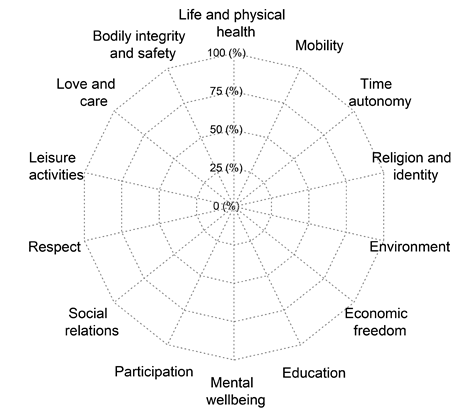}
\label{fig:spiderweb}
\end{figure}

Figure \ref{fig:spiderweb} shows the spiderweb chart designed to visualize the 14 dimensions of Multidimensional Child Growth (MICG). This framework, grounded in the work of \citet{Biggeri2006}, is informed by the Capability Approach and aligns closely with the CRC. Additionally, the MICG framework aligns with international policy agendas such as the UN Sustainable Development Goals (SDGs), particularly SDG~3 (Good Health and Well-being), SDG~4 (Quality Education), SDG~10 (Reduced Inequalities), and SDG~16 (Peace, Justice, and Strong Institutions). It provides an actionable bridge between high-level human rights commitments and concrete, measurable outcomes that reflect child growth and the context in which a child thrives. 

\begin{figure}
\caption{Chart of anthropometric child growth from the WHO standards for monitoring physical child growth. Source: WHO child growth standards (WHO, 2006, Figure 95, p. 223).}
\medskip
\centering
\includegraphics[width=0.7\textwidth]{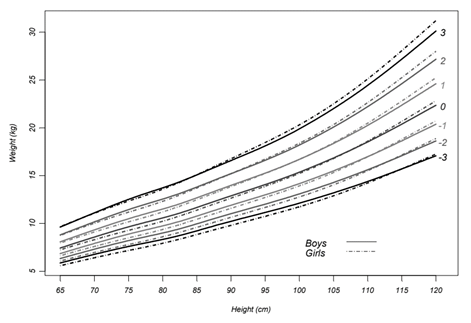}
\label{fig:who}
\end{figure}

We argue that this visualization diagram can be used for monitoring multidimensional child growth at both national and individual levels, in a similar way that WHO growth charts are used for monitoring the weight and height of children (Figure \ref{fig:who}). The axes of the chart of multidimensional growth measure the percentage of achievements in each of the 14 dimensions of child growth, based on the indicators described in Table \ref{tab:micg_yls_indicators}. Each axis in the spiderweb corresponds to a core dimension of child growth: \textit{Life and Physical Health, Mobility, Time Autonomy, Religion and Identity, Environment, Economic Freedom, Education, Mental Wellbeing, Participation, Social Relations, Respect, Leisure Activities, Love and Care,} and \textit{Bodily Integrity and Safety}. These dimensions reflect not only the instrumental conditions for physical and non-physical child growth but also intrinsic, context-dependent environmental determinants of child growth. The values plotted along each axis are scaled from 0 to 100\%, representing the degree of achievement or access in each domain, thereby enabling a comparative, profile-based interpretation.

From an academic standpoint, this visualization addresses long-standing challenges in child development research, particularly the difficulty of integrating qualitative, experiential dimensions with measurable, policy-relevant indicators. The MICG framework moves beyond reductionist monitoring and evaluation of child growth based solely on anthropometrics and foregrounds children’s agency, subjectivity, and contextual embeddedness. It reflects an ontological shift from viewing children as passive recipients of goods and services to recognizing them as active social agents with their own evolving capabilities.

The spiderweb visualization also serves as a heuristic for identifying synergistic or compensatory relationships among dimensions. For instance, limitations in \textit{Economic Freedom} may co-exist with high \textit{Love and Care} or \textit{Social Relations}, which may buffer some of the adverse effects. Conversely, deficiencies in \textit{Respect} or \textit{Bodily Integrity and Safety} may have cross-cutting implications for other dimensions, such as mental health or participation. Thus, the framework invites complex systems thinking, highlighting feedback loops and threshold effects that traditional linear models may overlook. Moreover, the inclusion of dimensions such as \textit{Religion and Identity} and \textit{Time Autonomy} expands the epistemic boundaries of dominant development paradigms by incorporating cultural, emotional, and temporal freedoms, which are often marginalized in both academic metrics and global indices. Contextualization is needed to properly design questions and instruments that capture concepts such as love and care consistently across cultural contexts.

For policymakers working on health and nutrition interventions, the spiderweb offers a strategic tool for diagnostic assessment, resource allocation, and monitoring. It enables stakeholders to visualize areas of underperformance and strength across a range of well-being domains and to identify where integrated interventions might yield the greatest impact across multiple micro-, meso-, and macro-levels. For example, persistent deficits in \textit{Mobility} or \textit{Time Autonomy} may suggest the need for child-friendly urban planning or more flexible schooling schedules, whereas gaps in \textit{Mental Wellbeing} might point to the urgency of scaling up psychosocial support services in schools and communities. It provides an integrative lens that bridges academic inquiry and policy implementation, empowering stakeholders to pursue holistic, context-sensitive, and justice-oriented interventions in support of children’s flourishing.

Finally, a methodological concern when assessing multidimensional child growth is the focus on realized and observable outcomes—such as being vaccinated—rather than on the underlying potentialities central to the concept of capability. This limitation makes it difficult to capture the full range of freedoms and opportunities that children may or may not have access to, since the MICG measures the achievements of children based on observed indicators, with achievements being equal to the observed situation.

To address this limitation and to measure the opportunities available to children—that is, their potential child growth above or below their achievements—we applied Bayesian stochastic frontier analysis to the MICG following the approach of \citet{henderson2020bayesian}. The Bayesian model estimates potential alternative values of multidimensional child growth that could be negatively or positively affected by children’s observed characteristics. We used geographical location and sex as discriminating characteristics, in addition to random circumstances that affect the achievements measured by the MICG. Specific children that could be left behind are those with the lowest opportunities for physical and non-physical growth—that is, those with the lowest estimated values of potential MICG.

\begin{figure}
\caption{From outcomes to opportunities and risks for healthy child growth: Illustration of the Bayesian approach for multidimensional child growth}
\medskip
\centering
\includegraphics[width=0.8\textwidth]{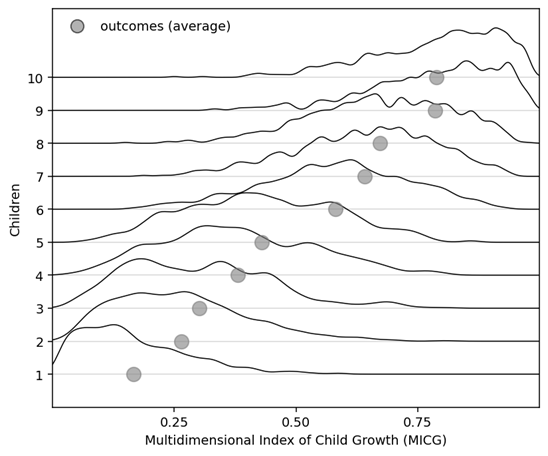}
\label{fig:micgbay}
\end{figure}

Figure \ref{fig:micgbay} illustrates the Bayesian extension of the Multidimensional Index of Child Growth (MICG), transitioning the focus from achievements (outcomes) to opportunities. The horizontal axis (MICG) represents the achievements of children based on their observed physical and non-physical conditions. These are direct measurements (e.g., health, cognitive scores, safety perception). Each curve in the vertical axis represents an individual child (in this case 10 children), and the probability density functions reflect their estimated opportunities for child growth, that is, the probability of achieving a specific outcome. Densities closer to a MICG equal to one reflect higher estimated opportunities for multidimensional growth, while density curves close to zero indicate lower opportunities to achieve desired outcomes. These opportunities (capabilities) are derived using Bayesian stochastic frontier analysis. The gray dot on each curve marks the actual multidimensional achievement of each child. It indicates where they currently stand in terms of growth. Compared to observed achievements, the Bayesian estimation of opportunities provides a counterfactual frontier that reflects the level of multidimensional growth each child could theoretically achieve under different circumstances. Based on the capabilities framework, the core idea of applying a Bayesian extension is that while two children might have the same achievements, their latent opportunities—their potential to grow physically, emotionally, and cognitively—may differ drastically. The Bayesian extension of the MICG thus allows the identification of children that could be “left behind” not just in actual outcomes, but in unrealized potential.

\FloatBarrier

\section{Proof-of-Concept: Macro-, Meso-, and Micro-Level Prototypes of the IUNS-MICG}\label{sec:applications}

\subsection{Macro-level: Using the MICG at population level for comparisons between countries}

We calculated a Multidimensional Index of Child Growth (MICG) to identify physical and non-physical deprivations experienced by 5-year-old children living in Ethiopia, Vietnam, India, and Peru. In the Young Lives Survey (YLS), the proportion of female and male children participating was similar across Ethiopia, Peru, Vietnam, and Andhra Pradesh and Telangana in India (Table~\ref{tab:yls}), but the percentage of rural children was higher in Ethiopia (60\%), Andhra Pradesh and Telangana in India (74\%), and Vietnam (79\%), compared to Peru (31\%). 

\begin{table}[h!]
\centering
\caption{Frequency distribution of children in the Young Lives Survey, by sex and region (urban/rural). Percentages of total below each frequency}
\medskip
\label{tab:yls}
\begin{tabular}{lcccccccc}
\hline
\textbf{Country} & \multicolumn{2}{c}{\textbf{Urban}} & \multicolumn{2}{c}{\textbf{Rural}} & \multicolumn{2}{c}{\textbf{Total}} & \multicolumn{2}{c}{\textbf{Total}} \\
 & Male & Female & Male & Female & Urban & Rural & Male & Female \\
\hline
Ethiopia & 399 & 362 & 611 & 540 & 761 & 1151 & 1010 & 902 \\ 
 & (20.87) & (18.93) & (31.96) & (28.24) & (39.8) & (60.2) & (52.82) & (47.18) \\ \midrule
India & 271 & 226 & 765 & 684 & 497 & 1449 & 1036 & 910 \\
 & (13.93) & (11.61) & (39.31) & (35.15) & (25.54) & (74.46) & (53.24) & (46.76) \\ \midrule
Peru & 694 & 666 & 296 & 307 & 1360 & 603 & 990 & 973 \\
 & (35.35) & (33.93) & (15.08) & (15.64) & (69.28) & (30.72) & (50.43) & (49.57) \\ \midrule
Vietnam & 213 & 194 & 800 & 763 & 407 & 1563 & 1013 & 957 \\
 & (10.81) & (9.85) & (40.61) & (38.73) & (20.66) & (79.34) & (51.42) & (48.58) \\
\hline
\end{tabular}
\end{table}

\begin{figure}
\caption{Percentage of children with achievements in each dimension of child growth}
\medskip
\centering
\includegraphics[width=\textwidth]{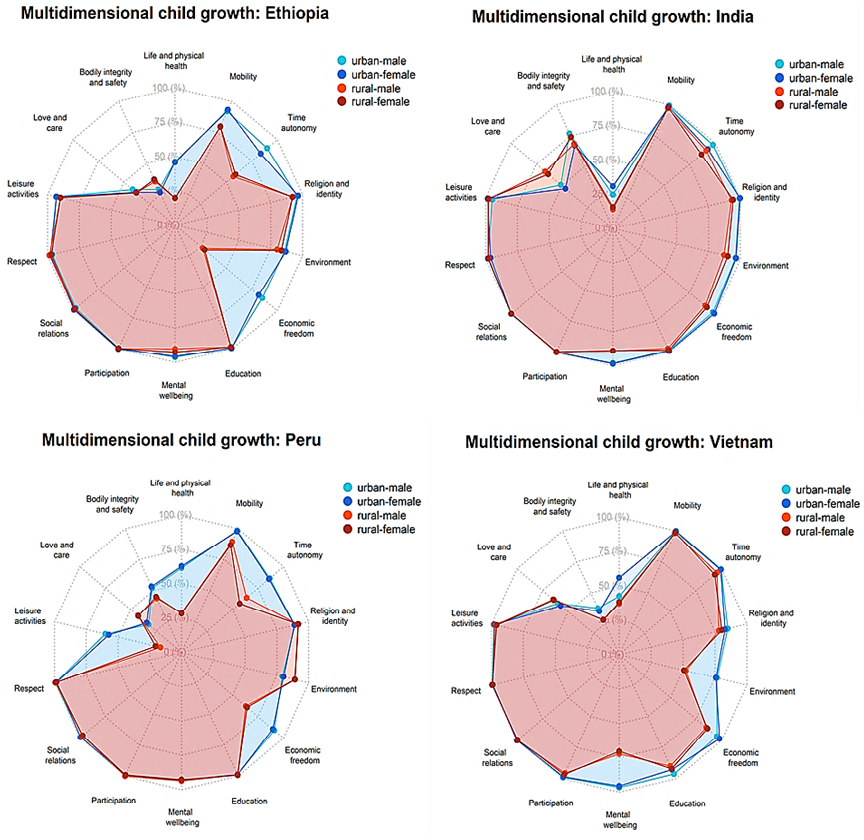}
\label{fig:micg4}
\end{figure}

Figure \ref{fig:micg4} shows the multidimensional achievements of children in the 14 dimensions of physical and non-physical growth. Compared to their rural peers, children in urban areas of Ethiopia and Peru had more economic freedom from paid and unpaid work and more time autonomy from domestic tasks, as well as improved physical health. In Vietnam, urban children had higher achievements in cognitive and verbal development compared to rural children. 

When comparing countries, children in Peru and Ethiopia were better off than their peers in Vietnam and India (Andhra Pradesh and Telangana) in the dimensions related to responsibility, obedience, and respect for elders, social relations with friends, mental wellbeing, and participation and cooperation at home. However, they were deprived of safety at home and in the streets, which can indirectly affect their survival. Compared to the other countries, children in Andhra Pradesh and Telangana in India were the most deprived in the life and physical health dimension, which includes anthropometric, nutrition, and health indicators.

In Peru, rural girls (red dots) experience the most acute deprivations across nearly all dimensions, particularly in education, mental wellbeing, economic freedom, and mobility. Their MICG profile shows the most contracted pattern, indicating a pervasive lack of opportunities and support.

Rural boys (red triangles) show slightly better outcomes than rural girls but still fall significantly behind their urban peers. Notably, they also experience severe deficits in participation, social relations, and bodily integrity and safety. In contrast, urban boys (light blue circles) perform better across nearly all indicators, particularly in mobility, time autonomy, and education. Their overall MICG profile is more expanded, suggesting greater access to enabling environments and institutional support. Urban girls (light blue squares) show relatively high achievements in mental wellbeing, participation, and respect, though their scores in economic freedom and bodily integrity suggest that gender-based constraints persist even in urban settings.

These multidimensional patterns point to complex interactions between structural factors—such as urban infrastructure, access to education, and economic opportunities—and cultural norms related to gender and childhood. The markedly low levels of participation, respect, and leisure activities for rural children indicate that children in these contexts may not only be materially deprived but also excluded from social and civic life, with potential long-term consequences for their agency and development.

From an academic perspective, these results underscore the added value of multidimensional measures over traditional unidimensional indicators (e.g., stunting or income poverty). The MICG captures invisible or underrepresented aspects of deprivation, such as lack of autonomy, poor mental health, or absence of respect, which are vital for understanding children’s actual freedoms and capabilities.

Policy-wise, these findings demand targeted, intersectional interventions. For instance, addressing rural girls’ deficits requires more than just increasing school enrollment and calls for integrated programs that also promote mental health, bodily safety, and social inclusion. Likewise, policies that aim to reduce rural–urban gaps must be sensitive to gender-specific vulnerabilities, ensuring that improvements in infrastructure or service delivery are inclusive. In summary, the MICG framework reveals that in Peru—as in other countries—rurality and gender intersect to shape distinct patterns of child disadvantage. Only by recognizing these multidimensional, overlapping forms of deprivation can we move toward more just, inclusive, and effective policies for childhood development.

This kind of evidence can inform child-centered planning in alignment with the Sustainable Development Goals (especially SDG~4 on education, SDG~5 on gender equality, and SDG~10 on reducing inequalities). Moreover, the spiderweb chart serves as a powerful communicative tool for stakeholders, policymakers, NGOs, and communities, as it can be used to visualize where deprivations are most acute and where resources should be prioritized.

\subsection{Meso-level: Improving the understanding of the impact of socio-economic policies at sub-national level}

A multidimensional index of child growth can also be used to improve the understanding of the impact of socio-economic policies at community level. In \citet{gonmart2022}, we investigate the role of community participation in child development. Drawing from the Vietnam Young Lives Study, we explore how collective engagement in child-rearing and nutrition programs enhances multidimensional child growth. Our findings indicate that when communities are actively involved in health and nutrition initiatives, children benefit from better health outcomes, social support, and improved access to resources.

\begin{figure}
\caption{(A–D) Effects of community participation on multidimensional child growth: WASH intervention—design stage. (A) Marginal effects of community participation on multidimensional child growth calculated with the ordinary least squares regression. Panels B, C, and D show the fit of a kernel density to the distribution of MICG in rural and urban communities. The dotted lines represent the median of the empirical distribution. MICG, Multidimensional Index of Child Growth; WASH, water, sanitation, and hygiene. Source: Gonzales Martinez, R., Wells, J., Anand, P., Pelto, G. Dhansay, M.A., Haisma, H. Community participation and multi-dimensional child growth: evidence from the Vietnam Young Lives Study. Current developments in nutrition (6) 4, p. nzac022 \citep{gonmart2022}}
\centering
\includegraphics[width=\textwidth]{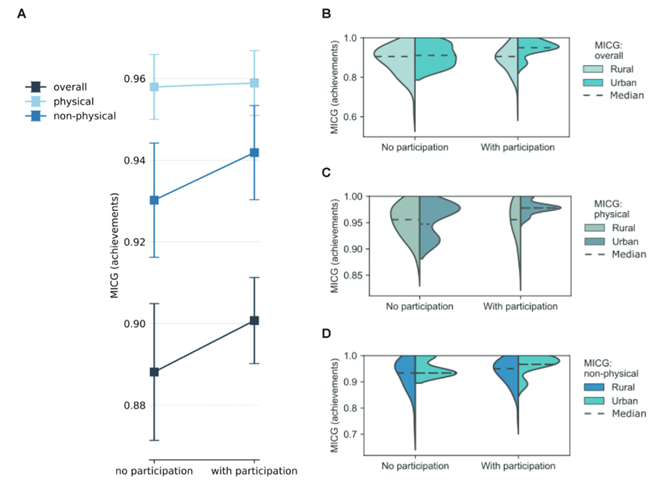}
\label{fig:comm}
\end{figure}

This study was the first to develop and apply the Multidimensional Index of Child Growth (MICG), grounded in the capability approach and constructed from 14 dimensions and 29 indicators spanning both physical (e.g., stunting, food diversity) and non-physical (e.g., cognitive development, emotional care, autonomy) dimensions, as discussed previously. Using ordinary least squares and quantile regression models, we found that community participation during the design stage of water, sanitation, and hygiene (WASH) interventions was significantly associated with improvements in multidimensional child growth, particularly among the most deprived children. In contrast, participation in the implementation stage showed no significant effect, and participation in the design of health interventions was negatively associated with growth, possibly reflecting mismatches between institutional goals and community needs (Figure \ref{fig:comm}).

Importantly, the benefits of participation were concentrated in the non-physical dimensions of child growth—such as psychological well-being and social support—underscoring the inadequacy of using only anthropometric indicators in program evaluations. Moreover, children in urban communities experienced more substantial gains from participatory interventions than those in rural areas, suggesting that local infrastructure and governance capacity may mediate the effectiveness of community involvement. Our study underscores the importance of fostering inclusive, context-sensitive participatory approaches during the early planning phases of public health strategies. By capturing the multidimensional nature of child well-being, such approaches not only enhance intervention effectiveness but also contribute to reducing inequality and promoting sustainable, community-driven development.

\subsection{Micro-level: MICG as a tool for dialogue that can be used to identify children that could be left behind by interventions}

The MICG and its visualization through a spiderweb chart offer a multidimensional lens to interpret developmental risks beyond anthropometric measures, enabling a more holistic understanding of child growth as a composite of physical, cognitive, and emotional health within the child’s broader context. The spiderweb chart is particularly useful as a dialogue tool between parents or caregivers and health professionals, as it supports more targeted referrals, anticipatory guidance, and collaborative care planning with educators and social service providers. Spiderwebs can also be used diagnostically to track the evolution of children’s achievements across multiple dimensions over time. For example, dynamic spiderwebs allow health professionals to visualize changes in achievements across dimensions and interpret a child’s “growth” at a certain point in time—such as during challenging life events—as well as across developmental trajectories.

The policy and clinical utility of the MICG at the micro-level is further enhanced by its communicative clarity: it can be used not only for technical reporting but also for participatory planning processes involving children, caregivers, and community stakeholders. By making inequalities and multidimensional deprivation visible in an intuitive format, the MICG and its spiderweb visualization contribute to evidence-informed, rights-based policymaking that is sensitive to both quantitative outcomes and qualitative life conditions. The MICG operationalizes a child-centered, capability-based, and rights-aligned approach to human development. It underscores the fundamental principle that child development must be understood not as a singular trajectory but as a multidimensional unfolding shaped by intersecting freedoms, resources, relationships, and environments.

\begin{figure}
\caption{Peru: (a) MICG (children’s achievements) and (b) the risk of being left behind by programs/interventions, calculated as lower opportunities of child growth. Each dot represents a child. Children with the highest risk of being left behind during the development are highlighted in a purple box in figure (b).}
\centering
\includegraphics[width=0.9\textwidth]{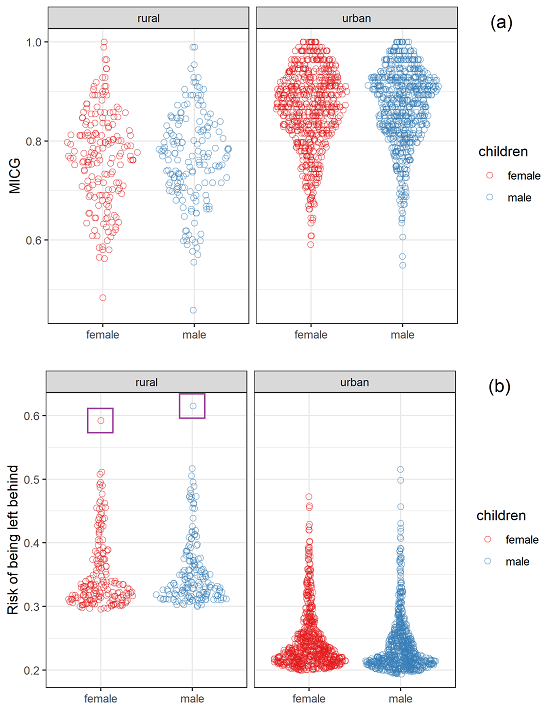}
\label{fig:LNBperu}
\end{figure}

\begin{figure}
\caption{India: (a) MICG (children’s achievements) and (b) the risk of being left behind by programs/interventions, calculated as lower opportunities of child growth. Each dot represents a child. Children with the highest risk of being left behind during the development are highlighted in a purple box in figure (b).}
\centering
\includegraphics[width=0.9\textwidth]{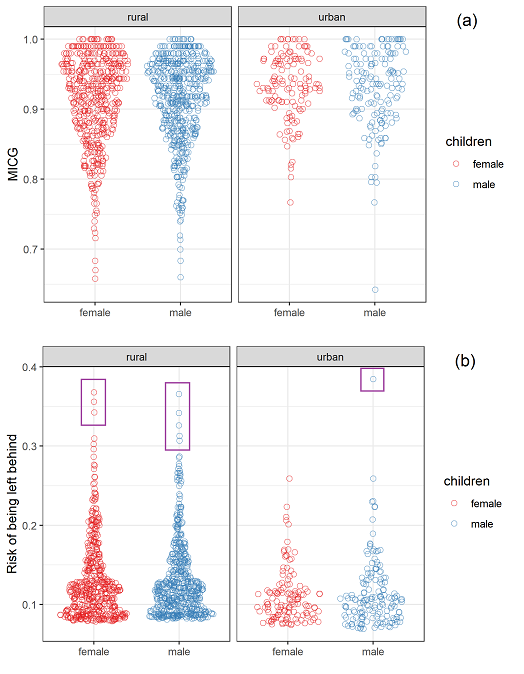}
\label{fig:LNBindia}
\end{figure}

\begin{figure}
\caption{Vietnam: (a) MICG (children’s achievements) and (b) the risk of being left behind by programs/interventions, calculated as lower opportunities of child growth. Each dot represents a child. Children with the highest risk of being left behind during the development are highlighted in a purple box in figure (b).}
\centering
\includegraphics[width=0.9\textwidth]{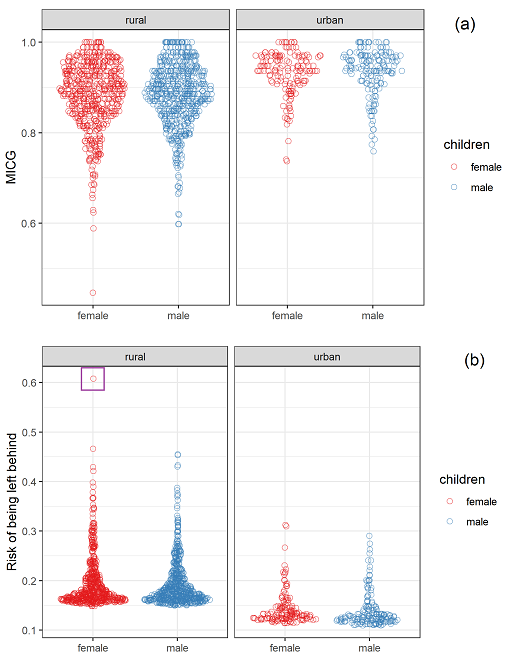}
\label{fig:LNBvietnam}
\end{figure}

\begin{figure}
\caption{Ethiopia: (a) MICG (children’s achievements) and (b) the risk of being left behind by programs/interventions, calculated as lower opportunities of child growth. Each dot represents a child. Children with the highest risk of being left behind during the development are highlighted in a purple box in figure (b).}
\centering
\includegraphics[width=0.9\textwidth]{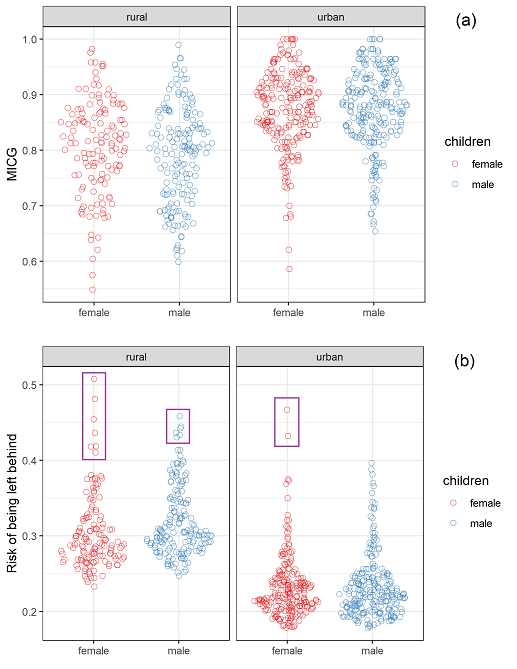}
\label{fig:LNBeth}
\end{figure}

In terms of identifying individual children who might be left behind by interventions, Figures \ref{fig:LNBperu}, \ref{fig:LNBindia}, \ref{fig:LNBvietnam} and \ref{fig:LNBeth} show the results of estimating capabilities with Bayesian methods for Peru, India, Vietnam, and Ethiopia. In the case of Peru, two children are highlighted for illustration: a rural girl and a rural boy, both displaying low opportunity estimates and at risk of being left behind in their multidimensional growth. Both children are exposed to structural and environmental vulnerabilities: they feel unsafe commuting to school, belong to minority ethnic groups, and live in areas prone to natural hazards. While the girl is stunted and scores low in both cognitive and verbal assessments, she is sociable and lives far from school, suggesting mobility deprivation. The boy, who is not physically stunted, struggles socially, has a low cognitive score, but lives closer to school and thus faces fewer mobility constraints. Although their measured achievements might not differ dramatically, their opportunity distributions do—exposing subtle yet critical disparities in their developmental trajectories. The visual separation in the figure, marked by the different colors for male and female children across rural and urban contexts, underscores how these inequalities interact with geography and gender. By estimating potential growth capabilities with Bayesian methods, our analysis offers a powerful diagnostic for targeting policies where they are most needed, revealing that equality in outcomes may mask deep-rooted inequalities in opportunities.

\FloatBarrier

\section{Conclusions}\label{sec:conclusions}

To further reduce child mortality and its inequalities, the IUNS-MICG Task force developed a capability approach for child growth that goes beyond anthropometry and includes non-physical dimensions that promote multidimensional child growth, like the care received from parents, cognitive and verbal development, and children’s safety. In previous studies \citep[e.g.][]{Haisma2018, gonmart2022}, we conceptualized, operationalized, and showed applications of the multidimensional index of child growth (MICG), which can be used for policy formulation/evaluation and for clinical practice. the theoretical framework using data from the YLS. In line with studies from the WHO–UNICEF–Lancet Commission \citep{Clark2020}, our key empirical results highlight the importance of having data sets that capture multiple dimensions of physical and non-physical child growth. For children that are stunted or wasted, our approach provides insights into the potential causes of malnutrition that remain hidden if the environmental context of parents and households is not considered. Our multidimensional approach, as well, helps to understand development deprivations of children that are not malnourished. 

Compared to previous child indicators at national level, our methodology is focused on the situation and opportunities of individual children, but at the same time our method allows us to compare results between countries. While child indicators based on large datasets—like the Multiple Indicator Cluster Surveys (MICS) and the Demographic and Health Surveys (DHS)—increase representativeness and facilitate cross-country comparisons, our findings show that suitable contextual and multi-dimensional information is needed to uncover social and psychological deprivations, which can help to further reduce child malnutrition and mortality, because complex multidimensional indicators at child level inform context-specific programs and focalized interventions for health-care professionals and policy makers—governments, international donors, and non-governmental organizations—, in line with the goals of the SDGs. 

As part of the operationalization of the MICG, we also proposed a visualization diagram for monitoring multidimensional child growth that can be potentially applied to complement the traditional growth charts of weight and length/height by age and sex used by pediatricians worldwide. Our visualization diagram of multidimensional child growth can be used for monitoring the social and psychological (verbal and cognitive) growth of children, as well as the influence of the environment on their physical growth. 

Because the purpose of the IUNS-MICG project was to create a prototype of MICG that can serve as a proof-of-concept of methodologies to measure opportunities for child growth, the results were not aimed to be generalizable at country level and are not directly comparable to those obtained from other surveys, like DHS and MICS, because the YL survey is based on purposive sampling with a pro-poor bias aimed at maximizing regional and socio-economic diversity. Future research focused on expanding the IUNS-MICG approach to multidimensional child growth needs to have richer datasets, that is, contextualized datasets that allow the calculation of indicators to cover the multiple dimensions of physical and non-physical child growth. These richer datasets, however, are currently not available, to the best of our knowledge. 

Future research can also incorporate how a nurturing environment and social trajectories inform opportunities for multidimensional child growth, based on a geometric-topological reformulation of Bronfenbrenner’s bioecological model using non-Euclidean differential geometry where the nested ecological systems (micro-, meso-, exo-, macro-, and chronosystem) are no longer idealized as concentric Euclidean spheres but as hyperboloidal manifolds embedded in a curved, anisotropic space. In Appendix B at the end of this document, this potential future research line is explained in detail.

\bibliographystyle{apalike}
\bibliography{references}

\newpage

\section*{Appendix A: The Determinants of Child Growth among the Worst-Off with a Multidimensional Framework }

\citet{Klennert2025} addressed the limitations of using traditional biomedical indicators—such as stunting and wasting—to assess child growth, particularly among the most deprived children. Recognizing the complexity of the double burden of malnutrition (DBM), where undernutrition coexists with increasing rates of childhood overweight, the study argues for a multidimensional conceptualization of child growth rooted in the capability approach. The research identifies a critical gap in current policy and intervention design: most studies focus on average effects or use single-dimensional measures, thus failing to capture the nuanced deprivation experienced by the worst-off children. To bridge this gap, \citet{Klennert2025} uses a dual-dataset approach, drawing from the Peruvian Demographic and Health Survey (DHS) and the Young Lives Study (YLS) for Peru. While the DHS offers broader coverage with limited contextual variables, the YLS provides rich, child-centered data with extensive information on social context and caregiving environments.
A central methodological feature of the study is the use of quantile regression at the 10th percentile (with robustness checks at the 15th) to focus specifically on the lower tail of the growth outcome distribution, effectively isolating the experience of the worst-off children. Two outcome variables are compared: a traditional biomedical indicator (stunting z-scores) and a Multidimensional Index of Child Growth (MICG).	
The findings show that using different frameworks yields substantially different profiles of deprivation. For instance, in the DHS data, indigenous ethnicity, maternal BMI, employment, and wealth are significant predictors of stunting. However, when MICG is used, gender becomes a significant factor (with girls faring better), and maternal employment or wealth lose their explanatory power. Similarly, in the YLS data, male rural indigenous children consistently emerge as the most disadvantaged under both frameworks, but the determinants differ: stunting is associated with rural residence and household size, whereas MICG highlights maternal BMI as a crucial protective factor, emphasizing the role of maternal health in multidimensional development.
The report highlights that data limitations in the DHS constrain the ability to fully implement a multidimensional framework, while the YLS lacks depth in contextual covariates. These contrasts reveal the trade-offs between breadth and depth in dataset selection and underline the need for integrated surveys that capture both determinants and outcomes comprehensively. The study makes a strong case for moving beyond average-based and biomedical-only assessments of child growth. It shows that a multidimensional framework is not only theoretically richer but also yields different empirical insights—identifying different children as the worst-off and pointing to different intervention targets. The use of quantile regression proves particularly valuable for policy-relevant research, enabling tailored responses for the most vulnerable populations. The study encourages future work to refine multidimensional indices, improve data infrastructure, and consider the contextual nature of growth to ensure that the most deprived children are not overlooked in policy design.

\newpage

\section*{Appendix B: Non-Euclidean Extension of Bronfenbrenner’s Bioecological Theory}

Bronfenbrenner’s bioecological theory conceptualizes human development as the result of dynamic interactions between multiple nested and interdependent environmental systems: the \textit{microsystem}, \textit{mesosystem}, \textit{exosystem}, \textit{macrosystem}, and \textit{chronosystem} \citep{Bronfenbrenner1977,Bronfenbrenner1979,BronfenbrennerMorris1998,Bronfenbrenner2005}. Traditionally, \textbf{these layers are depicted as concentric Euclidean circles/spheres}, implying metric independence and orthogonality among dimensions. Yet, as \citet{Bronfenbrenner2005} emphasized, real developmental processes are non-linear, context-dependent, and historically contingent. Feedback among social, structural, and temporal layers introduces curvature, anisotropy, and deformation that cannot be represented within Euclidean geometry.

To capture these interdependencies, the ecological space can be reformulated as a \textit{non-Euclidean manifold}, where curvature encodes the intensity and directionality of social and environmental influence. In this framework, each ecological level is represented as a hyperboloidal surface embedded in a pseudo-Riemannian space. The macrosystem bends the mesosystem and microsystem, while the chronosystem dynamically deforms the manifold over time, producing a continuously evolving geometry of development.

Let $(x, y, z)$ denote local ecological coordinates corresponding to micro-, meso-, and macro-level domains. The local metric tensor is defined as
\[
ds^{2} = dx^{2} + dy^{2} - dz^{2},
\]
a Lorentzian form that allows expansion and contraction along structural or temporal gradients. The negative term reflects the potential for distal structural forces to fold or invert local developmental surfaces—policy shifts, economic crises, or cultural transformations can thus be interpreted as geometric distortions propagating across ecological layers.

The manifold can be embedded in hyperbolic coordinates:
\[
x = \sinh(r)\sin(v)\cos(u), \qquad
y = \sinh(r)\sin(v)\sin(u), \qquad
z = \cosh(r),
\]
where $r$ denotes the ecological radius (distance from the developing person), $u$ an azimuthal angle (contextual direction), and $v$ a polar angle (temporal or relational inclination). The hyperbolic distance $r$ captures exponential amplification: small perturbations in the microsystem can propagate non-linearly through mesosystemic linkages and macrosystemic structures, consistent with Bronfenbrenner’s notion of \textit{proximal processes} that are reciprocally influential across levels \citep{BronfenbrennerMorris1998}.

In the classical Euclidean representation, the subsystems are orthogonal:
\[
f_x \perp f_y \perp f_z.
\]
In the curved ecology, however, these dimensions are dynamically coupled through curvature. Their interactions can be expressed by a rotationally coupled dynamical system:
\[
\begin{cases}
\dfrac{df_x}{dt} = \Phi_1(f_y - f_x),\\[6pt]
\dfrac{df_y}{dt} = f_x(\Phi_2 - f_z) - f_y,\\[6pt]
\dfrac{df_z}{dt} = f_x f_y - \Phi_3 f_z,
\end{cases}
\]
where $f_x, f_y, f_z$ represent evolving developmental potentials associated with the micro-, meso-, and macro-systems, and $\Phi_i$ are curvature coefficients determining the strength and direction of inter-level feedbacks. This non-linear system can exhibit attractors, bifurcations, and hysteresis—phenomena consistent with developmental discontinuities such as critical transitions, resilience thresholds, or sudden behavioral reorganizations.

Each system is described by a vector field $\bm{\Psi}_i$ representing proximal processes (e.g., child–caregiver interaction, educational influence, cultural exposure), modulated by curvature tensors $\bm{\Lambda}_i$ defining the strength of cross-level coupling:
\[
\bm{\Sigma} = \sum_i \bm{\Lambda}_i \bm{\Psi}_i.
\]
The chronosystem introduces time-dependent curvature:
\[
\frac{d\bm{\Psi}}{dt} = \bm{\Phi}(\bm{f}, t) + \kappa(t)\,\bm{\Psi},
\]
where $\kappa(t)$ represents the rate of temporal deformation due to historical change, migration, climate shocks, or technological transitions. This term operationalizes Bronfenbrenner’s \textit{chronosystem} as a differential operator acting on time, transforming static developmental ecology into a dynamic spatiotemporal manifold \citep{Bronfenbrenner2005}.

The trajectory of the individual within this curved ecology, $\mathcal{C}(t)$, is a geodesic minimizing ecological action:
\[
\mathcal{S} = \int \sqrt{g_{ij}\, \dot{x}^i \dot{x}^j}\, dt,
\qquad
\frac{d^2 x^k}{dt^2} + \Gamma^k_{ij} \frac{dx^i}{dt}\frac{dx^j}{dt} = 0,
\]
where $g_{ij}$ is the non-Euclidean metric tensor and $\Gamma^k_{ij}$ the Christoffel symbols encoding cross-level influences. The child’s developmental path bends in response to environmental curvature and torsion—an elegant mathematical analogue of Bronfenbrenner’s assertion that context and time jointly shape the course of development \citep{Bronfenbrenner1979}.

Within this geometric framework:
\begin{itemize}
  \item The \textbf{microsystem} corresponds to the local tangent plane—approximately Euclidean, dominated by immediate family, peers, and school environments.
  \item The \textbf{mesosystem} introduces torsion, representing the twisting and coupling of relationships and processes across immediate contexts.
  \item The \textbf{exosystem} induces anisotropy, distorting local curvature through indirect forces such as parental employment, institutional policy, or neighborhood conditions.
  \item The \textbf{macrosystem} defines the global curvature of the manifold, establishing large-scale social, cultural, and economic constraints that influence all subordinate levels.
  \item The \textbf{chronosystem} constitutes a time-dependent deformation field, where $\partial g_{ij}/\partial t \neq 0$, encoding historical transitions, life-course events, and temporal instability.
\end{itemize}

\begin{figure}
\caption{Non-Euclidean Extension of Bronfenbrenner’s Bioecological Theory. The hyperbolic surface represents the curved ecological manifold of human development. Concentric ecological rings (micro–meso–exo–macro) lie on the same curved topology, illustrating dynamic coupling between levels. Arrows (chronosystem) indicate temporal deformation of the manifold, while the bold curve shows an illustrative geodesic representing an individual’s developmental trajectory through time.}
\centering
\includegraphics[width=0.8\textwidth]{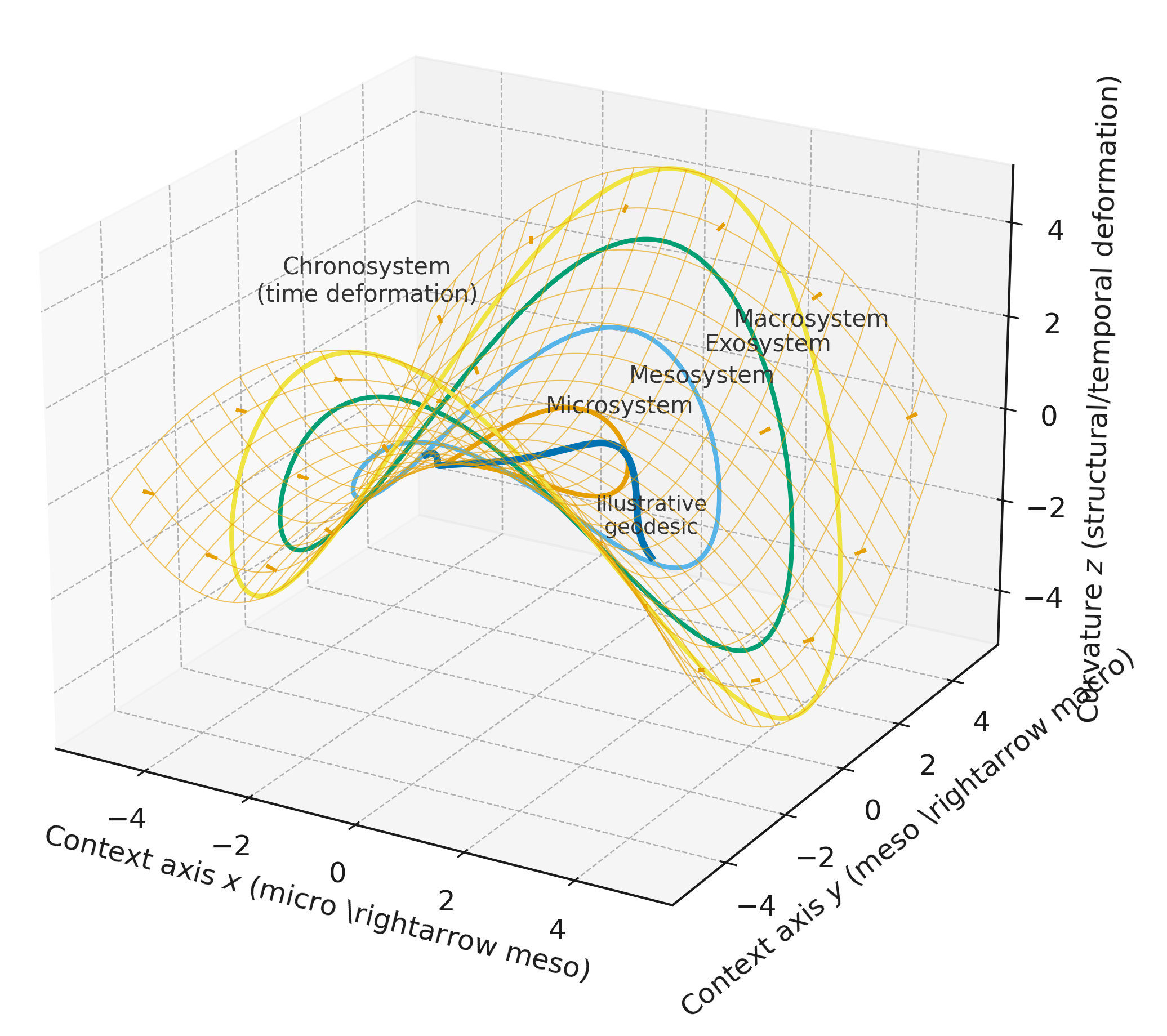}
\label{fig:bron}
\end{figure}

The ecological model thus becomes a \textit{hyperbolic developmental manifold}. Growth, resilience, and vulnerability correspond to geodesic accessibility across a curved landscape of interdependent systems. The resulting equation summarizing the ecological potential field is:
\[
\boxed{
\mathcal{E}(x, y, z, t) =
\sum_{i=1}^{n} 
\Lambda_i(t)\, \Psi_i(x, y, z)
\quad \text{subject to} \quad
ds^2 = dx^2 + dy^2 - dz^2.
}
\]
Here, $\mathcal{E}$ represents the ecological potential, $\Psi_i$ the directional influences of each system, and $\Lambda_i(t)$ their time-dependent curvatures. The framework transforms Bronfenbrenner’s bioecological theory from a descriptive ecological metaphor into a mathematically coherent topology of human development—one in which inequality, adaptation, and transformation emerge as the geometry of life trajectories within a curved, evolving social universe \citep{Bronfenbrenner1977,BronfenbrennerMorris1998,Bronfenbrenner2005}.

Figure \ref{fig:bron} shows the geometric reinterpretation of Bronfenbrenner’s bioecological model within a non-Euclidean (hyperbolic) developmental space. Instead of depicting nested concentric circles—as in the classical representation—the ecological systems are embedded in a curved manifold whose surface encodes the strength, direction, and distortion of environmental influence.

The saddle-shaped surface (a hyperbolic paraboloid) represents the pseudo-Riemannian manifold of development, characterized by both expansion and contraction along different contextual axes. This curvature conveys that developmental forces are not orthogonal or independent, but dynamically coupled: changes in distal social structures (macrosystem) can fold or distort local developmental contexts (microsystem). The curvature of the manifold expresses non-linearity and feedback among systems. Unlike the traditional model, where each layer acts independently, this visualization shows that forces from the macrosystem can “curve” the microsystem—an effect observable, for example, when national policy shifts reshape family routines or educational opportunities. The chronosystem adds a fourth dimension—time as curvature—transforming the static nested model into a living, morphogenetic space. Each historical or biographical change modifies the local geometry, continuously updating the developmental trajectory.

\end{document}